\documentclass[conference]{IEEEtran}
\IEEEoverridecommandlockouts
\usepackage{cite}
\usepackage{amsmath,amssymb,amsfonts}
\usepackage{algorithmic}
\usepackage{graphicx}
\usepackage{caption}
\usepackage{textcomp}
\usepackage{xcolor}
\usepackage{svg}
\usepackage{afterpage}
\usepackage{booktabs}
\usepackage{float}
\usepackage{subfigure}
\usepackage{multirow}
\usepackage{todonotes}
\usepackage{soul}

\def\BibTeX{{\rm B\kern-.05em{\sc i\kern-.025em b}\kern-.08em
    T\kern-.1667em\lower.7ex\hbox{E}\kern-.125emX}}
\begin{document}

\title{A Lightweight Multi-Attack CAN Intrusion Detection System on Hybrid FPGAs}


\author{\IEEEauthorblockN{Shashwat Khandelwal \& Shanker Shreejith}
\IEEEauthorblockA{Department of Electronic \& Electrical Engineering,
Trinity College Dublin\\
Dublin, Ireland\\
Email: \{khandels, shankers\}@tcd.ie}}

\maketitle

\begin{abstract}
Rising connectivity in vehicles is enabling new capabilities like connected autonomous driving and advanced driver assistance systems (ADAS) for improving the safety and reliability of next-generation vehicles. 
This increased access to in-vehicle functions compromises critical capabilities that use legacy in-vehicle networks like Controller Area Network (CAN), which has no inherent security or authentication mechanism. 
Intrusion detection and mitigation approaches, particularly using machine learning models, have shown promising results in detecting multiple attack vectors in CAN through their ability to generalise to new vectors.
However, most deployments require dedicated computing units like GPUs to perform line-rate detection, consuming much higher power. 
In this paper, we present a lightweight multi-attack quantised machine learning model that is deployed using Xilinx's Deep Learning Processing Unit IP on a Zynq Ultrascale+ (XCZU3EG) FPGA, which is trained and validated using the public CAN Intrusion Detection dataset. 
The quantised model detects denial of service and fuzzing attacks with an accuracy of above 99\% and a false positive rate of 0.07\%, which are comparable to the state-of-the-art techniques in the literature.
The Intrusion Detection System (IDS) execution consumes just 2.0\,W with software tasks running on the ECU and achieves a 25\% reduction in per-message processing latency over the state-of-the-art implementations.
This deployment allows the ECU function to coexist with the IDS with minimal changes to the tasks, making it ideal for real-time IDS in in-vehicle systems.

\end{abstract}

\section{Introduction \& Related Works}
Most high-end vehicles today integrate over 50 electronic computing units (ECUs) interconnected through different network standards for incorporating safety-critical, comfort and automation capabilities in a cost and energy-efficient manner. 
Controller Area Network (CAN) and its variants continue to be the most widely used network protocol in automotive electric/electronic systems owing to their low cost, flexibility, and robustness. 
CAN is a broadcast protocol with implicit priority using the \emph{ID} field and support for multiple data rates (from 125\,Kbps to 1\,Mbps), allowing their use in critical and non-critical network segments in the vehicle~\cite{CanBosch,hartwich2012can}.
CAN (and most other automotive networks) do not have native support for message (sender) authentication or encryption; however, due to the physical localization of earlier vehicular systems and the design of (critical) applications as silos (paritioned into domains), integrating them over unprotected CAN networks was considered secure. 
With increasing connectivity enforcing bridged access between previously siloed domains, security gaps and vulnerabilities in these networks have been exposed further through numerous exploits~\cite{miller2015remote}.
For instance, passive eavesdropping attacks enabled through malicious code or aftermarket extensions can lead to a complete loss of user privacy~\cite{enev2016automobile}, while more enhanced attacks like Denial of Service (DoS) can fully disrupt normal vehicle functioning.

Intrusion detection systems (IDSs) for high-speed CAN networks address these challenges, often leveraging hardware accelerators to analyse and identify attacks like DoS in real-time~\cite{lin2012cyber}.
Early rule-based IDS schemes used inherent network and message properties like message frequency, voltage profiles of ECUs, or timing of remote request frames as features to detect abnormal operations~\cite{cho2016fingerprinting,cho2017viden,lee2017otids}. 
However, the rule-based approach limits their ability to scale and generalise towards new attack vectors, while also incurring incremental storage and computational overheads for new attack modes/vectors.
More recently, both classical and deep-learning based IDSs have proved extremely efficient in detecting CAN bus attacks~\cite{narayanan2015using, alshammari2018classification, yang2019tree, song2020vehicle, tariq2020cantransfer}. 
In~\cite{song2020vehicle}, the authors use a deep-convolutional neural network (CNN) based on inception net architecture as an IDS taking the CAN IDs as the input features to achieve an average detection accuracy of 99\% on both the DoS and Fuzzing attacks.
Alternative network architectures like Generational Adversarial Networks (GAN)~\cite{seo2018gids}, temporal convolutional networks with global attention ~\cite{cheng2022tcan} and a combination of CNN and long short-term memory (LSTM) as an unsupervised learning network~\cite{agrawal2022novelads} have been explored in the literature to improve detection accuracy, using only CAN IDs or complete data frame as inputs. 
Most such models rely on GPU acceleration to deploy the IDS at an ECU.
Dedicated IDS-ECUs have also been proposed, where stacked models are executed on a Raspberry Pi-3 device that acts as an IDS-ECU~\cite{yang2021mth}. 
An intrusion prevention ECU has also been proposed where an anomaly detection algorithm is executed on a Raspberry Pi device (acting as a standalone ECU) to detect fuzzing and spoofing attacks~\cite{de2021efficient}. 
This approach drops any suspected messages from the network by triggering an error response on the bus; however, this can cause a large number of critical messages to be affected in case of false positives or DoS attacks. 
In all the above cases, the scalable and generalisable nature of ML models are deployed through specialised accelerators (GPUs) or dedicated ECUs and can adapt to new threats post-deployment, making them more attractive than traditional rule-based methods. 

An alternative architecture approach is through the use of hybrid FPGAs as ECUs, allowing a machine learning accelerator to be closely coupled with the application(s) running on the ECU for a software-controlled hardware IDS. 
Hybrid FPGA devices like Zynq Ultrascale+ combine automotive-grade ARM processors and a large programmable fabric closely integrated on the same die, allowing custom machine learning accelerators to be integrated on the same die.
Prior research has explored FPGA-based ECUs that enable custom accelerators and specialised network capabilities to be tightly integrated as a system-on-chip module while maintaining full AUTOSAR compliance~\cite{fons2012fpga} and for the acceleration of complex algorithms like advanced
driving-assistance system (ADAS) using network interface extensions to augment ECU functionality and security~\cite{shreejith2018smart}.
Machine learning on hybrid FPGAs has also shown tremendous promise in applying deep learning in multiple applications through both vendor tool flows and academic efforts~\cite{xilinxvitis,gorbachev2019openvino,wang2019lutnet,umuroglu2017finn,soltani2019real,nakahara2018demonstration,mahajan2016tabla}.

In this paper, we present a lightweight deep-learning model for multi-attack (MA) intrusion detection and its deployment using a hybrid-FPGA based ECU architecture modelling a standard software-controlled accelerator framework in ECUs.  
Our deep-CNN model is quantised and integrated using Xilinx's Vitis-AI flow using a Deep-learning Processing Unit (DPU) accelerator IP which is controlled through Python APIs for our test cases~\cite{xilinxdpu}.
We use an Ultra96-V2 board as a prototype ECU platform which uses a ZynqMP ZU3EG device running Linux on the ARM cores with the PYNQ libraries.
Our model is trained to simultaneously detect DoS and fuzzing attacks using the public CAN attack dataset that is widely used in literature.
Our tests show that the tightly coupled IDS achieves comparable detection accuracy as the state-of-the-art detection accuracy with 25.08\% reduction in the per-message processing latency over state-of-the-art Raspberry Pi based approach while consuming just 2.0\,W in operation. 
We believe that the proposed IDS model and its deployment architecture will pave the way for seamless integration of generalised IDS in future vehicular network architectures.

\section{System Architecture}\label{sec:architecture}
\subsection{Hybrid-FPGA based IDS capable ECU}

\begin{figure}[t!]
    \centering
    \includegraphics[scale = 0.65]{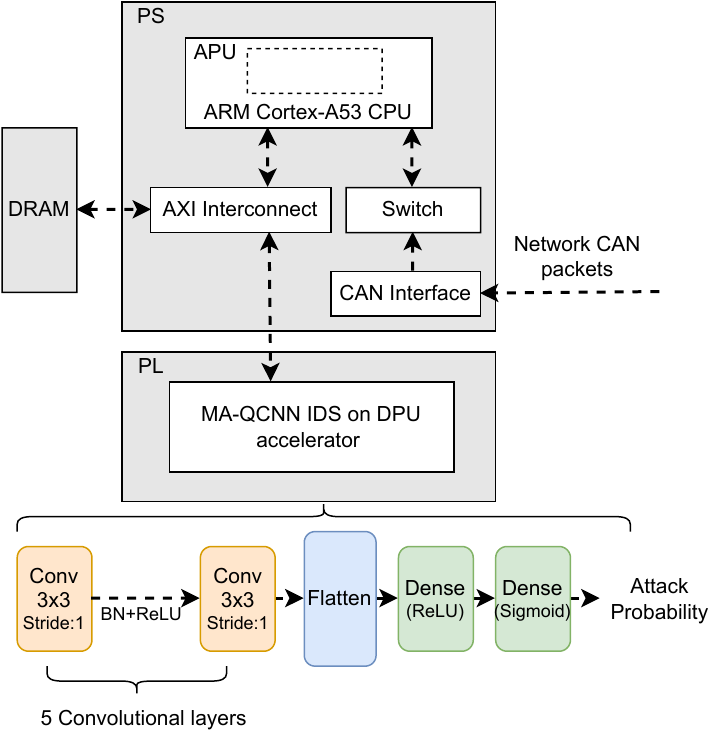}
    \caption{Proposed system architecture of the IDS-ECU. The ML model is accelerated on the PL part of the FPGA device.}
    \label{fig:arch}
    \vspace{-10pt}
\end{figure}

Figure~\ref{fig:arch} shows an overview of the proposed hybrid-FPGA based distributed ECU architecture on a Xilinx Zynq Ultrascale+ device.
The Zynq Ultrascale+ device integrates a quad-core ARM processor and a dual-core ARM real-time core within the processing system (PS) section of the device, along with integrated peripherals and interfaces.
The programmable logic (PL) section of the device enables the addition of specialised custom logic blocks or accelerators that can be accessed using the Advanced eXtensible Interface (AXI) protocol from the PS.
In this architecture, standard ECU function(s) are mapped as software tasks onto the ARM cores on the PS, on top of a standard operating system like linux or a real-time operating system. 
The operating system provides relevant drivers and APIs for accessing the PS peripherals and the PL accelerators, abstracting away low-level details of these blocks to create an AUTOSAR compliant architecture~\cite{fons2012fpga}.

In our approach, we propose to use the integrated CAN interface on the PS to handle the interfacing of ECU to the CAN bus, as shown in Figure~\ref{fig:arch}.
Any received packets at the interface are read by the software tasks on the PS and further processed based on the ECU's task specification.
In tandem, the IDS task extracts relevant packet feature information from the received packet and forwards it to the IDS accelerator through the integrated APIs.
For this evaluation, we have used the off-the-shelf DPU accelerator as the IDS inference engine, which executes our TensorFlow ML model. 
DPU is an instruction-based array of programmable processing engines (PEs) for accelerating deep-learning inference on FPGAs.
DPU is instantiated as a memory-mapped IP, by configuring the number of PEs and associated resources for configurable levels of parallelism based on resources available on the target FPGA.
The Vitis AI design flow compiles the TensorFlow model to executable instructions for the DPU's PE.
Standard Vitis AI Runtime (VART) APIs are used to configure and communicate with the DPU-based IDS engine.
The model uses interrupts to indicate the completion of tasks to the PS, allowing the software tasks to run in a non-blocking fashion.
This approach allows for seamless integration of the IDS API calls with the other event/time-triggered tasks to be executed by the ECU.
For high-performance networks like Automotive Ethernet, the IDS model can be configured to directly read the packets from the interface (also in PL), eliminating the bottleneck caused by the software-based data movement. 
In this case, the IDS task would enable the accelerator operation at the startup phase of the network and would be interrupted only when an anomaly is observed in the network traffic. 

\subsection{Dataset \& Training}
We use the CAR Hacking dataset for training and testing our model~\cite{song2020vehicle,canlink}.
The dataset provides a labelled set of normal and attack messages which were captured via the Onboard Diagnostic (OBD) port in an actual vehicle, with attack messages injected in real-time. 
The dataset includes generalised DoS and Fuzzing message injections (as well as other targeted attacks like gear-ECU spoofing) allowing us to validate the detection accuracy across these different attacks. 
DoS attacks flood the bus with a higher priority message when injected while a fuzzing attack transmits messages with random IDs and payloads at random intervals to disrupt the normal flow of messages. 
An extract from the dataset is shown in table~\ref{table:4}, showing the CAN ID (ID field), control field, data length code (DLC) and the actual data segment.
We ignored targeted attacks in the dataset since they are modelled by a constant message sequence (specific ID and payload) and are not generalisable to other ECUs.

As our input feature, we use a window approach to capture adjacent sequences of CAN messages, which are fed as input to the ML model to capture the correlation between messages under different operating conditions.
Through design space exploration, we observed that a timeseries window of 4 successive CAN-IDs provided optimal performance and was hence chosen for our implementation. 
The dataset is pre-processed to mimic byte-wide binary CAN data which can be read from the CAN interface controller.
The ID bits are extracted and stacked to form a block shape of \textit{\{4,11\}}, which is then time-spliced to generate the \textit{\{2,2,11\}} dimension that is fed as input to the model~\cite{tariq2020cantransfer}.
%
During implementation, this can be mapped as a FIFO style buffer stacking sequential CAN IDs from the CAN controller (in PL) to form the input tensor for the IDS.  

To train the model, we used the adam optimizer with the binary cross-entropy loss function.
The learning rate was set at 0.0001 during the entire training process to reduce any performance degradation when quantising the network post-training as observed in~\cite{wu2018training}.
We first train the model on DoS attack and ensure optimal performance; subsequently, this trained model is trained on the fuzzing dataset to improve generalisation across the two attack modes through inductive transfer. 
This model file is then tested in both the DoS \& Fuzzing attacks to ensure there was no performance degradation.
The model saves intermediate results at each epoch allowing us to progressively track and integrate early stopping in case of a significant drop in the accuracy.
The model was trained for 25 epochs with a batch size of 64 and the dataset was split as 80:15:5 for training, validation and testing respectively for both attacks to allow a sufficient amount of information for training and fine-tuning the quantised network. 
%

\begin{table}[t!]
\centering
\caption{An extract from the open Car hacking dataset which is used for our testing and evaluation.}
\scalebox{0.95}{
\begin{tabular}{@{}lrrr@{}}
\toprule
\textbf{Time}     & \textbf{ID} & \textbf{DLC} & \textbf{Data}           \\ \midrule
\textbf{$\ldots$} \\
1478198376.389427 & 0316            & 8            & 05,21,68,09,21,21,00,6f \\
1478198376.389636 & 018f            & 8            & fe,5b,00,00,00,3c,00,00 \\
1478198376.389864 & 0260            & 8            & 19,21,22,30,08,8e,6d,3a \\ 
\textbf{$\ldots$} \\
\bottomrule
\end{tabular}}
\label{table:4}\vspace{-10pt}
\end{table}

\subsection{Proposed Model}
To determine the model configuration with minimal computational complexity, we explored different network architectures that combine the layers supported by the Vitis-AI flow. 
Each model was defined in TensorFlow (TF) using standard TF functions and nodes. 
We evaluated different models comparing their inference latency and detection accuracy across both attack datasets when mapped using the Vitis-AI to determine our lightweight network model to implement. 
%
Our chosen model is composed of 5 \textit{Conv2D} layers implemented with 40, 80, 120, 160 \& 200 filters at each layer, each filter having a dimension of 3x3. 
This is followed by a \textit{Flatten} and 2 \textit{Dense} layers implemented with \{32, 1\} units respectively. 
The final dense layer consists of \textit{sigmoid} activation to predict the probability of an attack message.
In addition, batch normalisation and dropout layers were used between the convolutional layers to prevent overfitting and to improve learning efficiency during the inductive training phase.
\section{Deployment and Experimental Results}\label{exp_results}
Once the model was chosen, it was trained using an Nvidia RTX A6000 GPU using the training split of the dataset(s), starting with the DoS attacks and subsequently moving to the fuzzing attacks. 
The pre-quantisation accuracy is then verified using the test split of the dataset and is shown in Table~\ref{table:preqcomp}.
The exported model file is passed to the Vitis-AI tool flow which quantises the model, weights, biases, and activations to 8-bit precision, followed by an optional post-quantisation fine-tuning step to account for any performance degradation from the quantisation. 
In our case, the post-quantisation fine-tuned model performance showed no visible degradation in inference performance measuring using Precision, Recall, F1 score, False Negative Rates (FNR) \& False Positive Rate (FPR) metrics, as shown in Table~\ref{table:preqcomp}.
\begin{table}[t!]
\centering
\caption{Inference accuracy metrics of the proposed MA-QCNN model, pre and post quantisation on the two attacks.} 
\scalebox{0.95}{
\begin{tabular}{@{}lllllll@{}}
\toprule
\textbf{Attack} & \textbf{Model}  & \textbf{Precision} & \textbf{Recall} & \textbf{F1} & \textbf{FPR} & \textbf{FNR} \\
\midrule
\multirow{2}{*}{DoS} & Pre-Q          & 0.9992             & 1               & 0.9996    & 0.04\% & 0\%  \\
& MA-QCNN         & 0.9992             & 1               & 0.9996    & 0.04\% & 0\%  \\
\multirow{2}{*}{Fuzzy} & Pre-Q          & 0.9966             & 0.9875          & 0.992    & 0.1\% & 1.25\%  \\
& MA-QCNN         & 0.9966             & 0.9878          & 0.9922   & 0.1\% & 1.22\%   \\
\bottomrule
\end{tabular}}
\label{table:preqcomp}\vspace{-10pt}
\end{table}

The model is then packaged as an `xmodel' file which can be executed using a DPU accelerator in the PL with the VART libraries installed on the OS running on the PS. 
For our experiments, we use a petalinux based image for the Ultra-96 V2 board on which the PYNQ and VART libraries were installed to use the standard VART/PYNQ APIs. 
Subsequently, the bitstream corresponding to the DPU accelerator is loaded to the PL from the OS (as an overlay). 
For our experiments, we use the B1152 DPU for most of the performance measurements (unless specified) with the DPU using a 150 MHz interface clock and 300 MHz DSP core clock.

We compare our model performance in terms of inference accuracy metrics and processing latency with the state-of-the-art IDSs and IPSs described in the literature.
Performance metrics such as Precision, Recall, F1 score, and FNR are used to compare the inference accuracy of other models where reported and use per-message latency to capture the true latency of the detection approach starting from the arrival of a new CAN message at the interface. 
In case of schemes where inference is performed on a block of CAN messages, this is captured using the block size in the comparison.
We further determine the power consumption of the ECU while executing the IDS and compare it against the execution of the full-precision model on a GPU. 
Note that none of the ML-based IDS approaches in the literature report the actual power consumption of their implementations; however, since most models are significantly more complex than our lightweight model, we assume that our GPU implementation would be a good baseline for comparison.  

\subsection{Inference Accuracy}

Table~\ref{table:dcnncomp} compares the proposed MA-QCNN with the state-of-the-art approaches discussed in the research literature: GIDS~\cite{seo2018gids}, DCNN~\cite{song2020vehicle}, iForest~\cite{de2021efficient}, MTH-IDS~\cite{yang2021mth}, TCAN-IDS~\cite{cheng2022tcan} \& NovelADS~\cite{agrawal2022novelads}.
In the case of MTH-IDS, the authors only report an average accuracy, precision \& recall of 99\% across all attacks which is identical to our model on the three metrics.
In case of the \textit{DoS attack}, our quantised hardware accelerator performs almost identical to the DCNN, TCAN-IDS, NovelADS, MLIDS across all metrics and has superior performance than GIDS across all metrics.
In case of the \textit{fuzzing attack}, we observe that DCNN, MLIDS \& NovelADS perform marginally better, with 0.6\%, 0.7\% \& 0.8\% better F1 scores respectively. 
Our model is almost identical to TCAN-IDS in case of fuzzing attacks and offers a marginal improvement over iForest and GIDS by 1.7\% and 0.9\% respectively (F1 scores).
Note that in case of competing techniques, the models are trained afresh for each attack, whereas our model uses an induced learning approach during training and hence, uses the same weights, biases and activation values for both attacks.

\begin{figure}[t!]
    \centering
    \includegraphics[scale = 0.45,trim={0cm 1cm 0 1cm},clip]{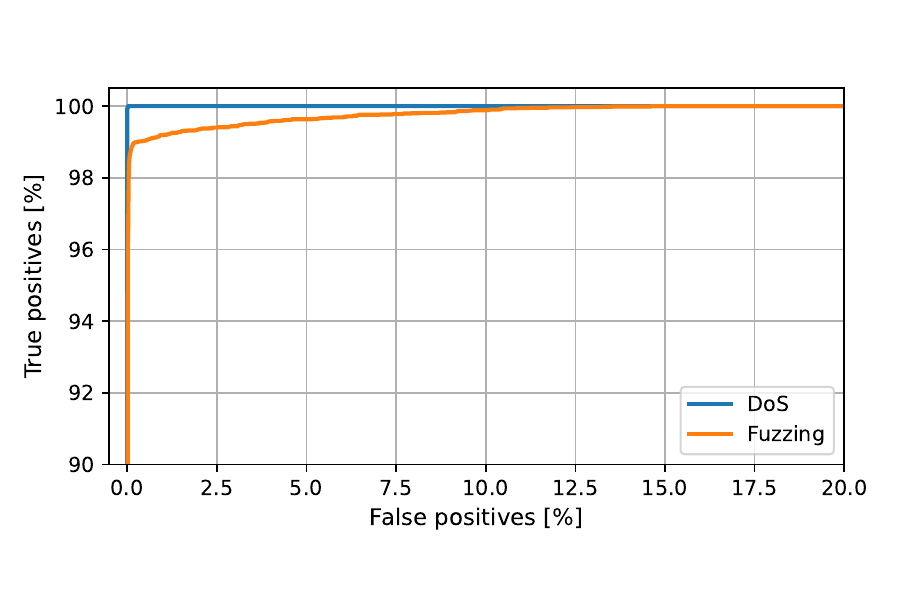}
    \caption{ROC curve of the MA-QCNN on the two attacks.}
    \label{fig:roc}
    \vspace{-15pt}
\end{figure}

Table~\ref{table:confmatrix} shows the confusion matrix of MA-QCNN on the two attacks. 
We also plot the receiver operating characteristic (ROC) curve for the two attacks when using our model on the Ultra-96 V2, which is shown in Figure~\ref{fig:roc}, and further determine the area under curve (AUC) values for our model based on the test datasets.
We observe that our model achieves an AUC value $ > $0.99 in both attacks with a slightly lower performance in the fuzzing attack.
We believe that this could be improved by fine-tuning the training and post-quantisation tuning steps.

\begin{table}[t!]
\centering
\caption{Confusion matrix capturing the classification results of the MA-QCNN (on DPU).}
    \scalebox{0.95}{
        \begin{tabular}{llrr}
            \toprule
            \textbf{Attack} & \textbf{Message Type}  & Predicted Normal  & Predicted Attack  \\ 
            \midrule
          \multirow{2}{*}{DoS} & True Normal    &  33282                                & 13                                   \\ 
            & True Attack    & 0                                 & 16705                                \\ 
           \midrule
            \multirow{2}{*}{Fuzzy} & True Normal    &    38784                            &   38                                 \\
           &  True Attack    &    136                              &   11042                              \\ 
            \bottomrule
        \end{tabular}}
\label{table:confmatrix}
\end{table}

\begin{table}[t!]
\centering
\caption{Accuracy metric comparison (\%) of our quantised FPGA accelerator (MA-QCNN) against the reported literature.}
\scalebox{0.95}{
\begin{tabular}{@{}llllll@{}}
\toprule
\textbf{Attack}  & \textbf{Model} & \textbf{Precision} & \textbf{Recall} & \textbf{F1}  & \textbf{FNR} \\
\midrule
\multirow{5}{*}{DoS} & GIDS~\cite{seo2018gids}                  & 96.8                & 99.6          & 98.1  &   -   \\ 
& DCNN~\cite{song2020vehicle}                  & 100                & 99.89          & 99.95  & 0.13     \\
& MLIDS~\cite{desta2020mlids}                  & 99.9                & 100          & 99.9  & -     \\
& NovelADS~\cite{agrawal2022novelads}                  & 99.97               & 99.91          & 99.94  & -     \\
& TCAN-IDS~\cite{cheng2022tcan}                  & 100             & 99.97          & 99.98  & -     \\
& iForest~\cite{de2021efficient}                  & -                &   -       &  - &  -    \\ 
& \textbf{MA-QCNN (DPU)}                  & 99.92             & 100               & 99.96  & 0     \\
\midrule
\multirow{5}{*}{Fuzzing} & GIDS~\cite{seo2018gids}                  & 97.3                & 99.5          & 98.3   & -    \\ 
& DCNN~\cite{song2020vehicle}                 & 99.95             & 99.65          & 99.80  & 0.5     \\
& MLIDS~\cite{desta2020mlids}                  & 99.9             & 99.9          & 99.9  & -     \\
& NovelADS~\cite{agrawal2022novelads}                  & 99.99               & 100         & 100  & -     \\
& TCAN-IDS~\cite{cheng2022tcan}                  & 99.96             & 99.89          & 99.22  & -     \\
& iForest~\cite{de2021efficient}                  & 95.07                & 99.93          & 97.44  &    -  \\
& \textbf{MA-QCNN (DPU)}            & 99.66             & 98.78          & 99.22   & 1.22     \\
\bottomrule
\end{tabular}}
\label{table:dcnncomp}\vspace{-10pt}
\end{table}

\subsection{Latency, Power consumption \& Resource utilisation}
We measure the cold-start setup time of IDS to quantify the time from power on to the model being ready. 
This is an important consideration since many attacks could be initiated during the startup phase of the network and ECUs.
In case of our test platform, we observe this cold-start latency as 8 seconds when averaged across multiple runs. 
It should be noted that our test platform uses a slower boot device (SD image) and full Linux OS, whereas typical ECUs rely on non-volatile flash memory and real-time operating system which could reduce the cold-start latency by up to 6$\times$.
%
Next,  we evaluate the processing latency of the inference datapath from the arrival of a new message and compare it against competing approaches in the literature. 
As mentioned before, some approaches perform block-based detection and hence require the block of messages to be available before it is fed as input to the inference engine.
This reduces average inference latency per message leveraging the parallelism on the GPUs.
However, accumulating 64 messages (for example) consumes nearly 8.3\,ms on a 1\,Mbps CAN network with 8-byte messages, which should be factored into their worst-case detection latency when averaging.
The results are shown in table~\ref{table:latcomp}, also capturing the execution platform and the block size in case of block-based inference. 
We observe the lightweight model on our approach takes 0.43\,ms from the arrival of a message from the CAN interface, when measuring the processing time to the arrival of the last message of the block, achieving a 1.3x speedup over state-of-the-art CAN IDSs.
Finally, our approach does incur additional latency due to message looping through the ECU, and could be reduced further using a CAN interface in the PL feeding the DPU through a FIFO. 
Also, on a more capable hybrid FPGA, a larger DPU could be configured to have multiple execution units and/or multiple execution threads to reduce latency further for higher speed CAN networks at the expense of slightly higher power consumption and device cost.

We also measure the power consumption of the IDS-ECU while performing inference, using the PYNQ-PMBus package to monitor the power rails directly. 
All power measurements are averaged over 100 inference runs, each evaluating the test set lasting nearly 10 seconds. 
We observe that the Ultra-96 ECU consumes 2\,W when the CAN messages are been analysed using the DPU block on the PL. 
In comparison, our lightweight floating-point model running on the A6000 consumes 54\,W while performing inference at similar CAN data rates, which is 27$
\times$ higher than the entire IDS-ECU on the Ultra-96 device. 
Other IDS implementations in the literature do not report power measurements; however, our lightweight model on the GPU should be a good baseline and more complex models discussed in the literature should consume higher power during inference on GPUs.
Table~\ref{table:resourceutilization} shows the resource utilisation on the PL, with the design consuming 45\% DSP blocks and 57\% BRAMs on the XCZU3EG device, leaving sufficient resources for other custom logic.
Thus, our IDS-ECU on a hybrid FPGA like an Ultra-96 device offers an ideal mix of detection accuracy, latency, power consumption, while offering software-driven control of IDS and seamless GPU-like deployment with minimal change to the ECU application, making it an ideal choice for in-vehicle deployment.

\begin{table}[t!]
\centering
\caption{Per-message latency comparison against state of the art IDSs (GPU/Raspberry Pi) reported in literature.}
\scalebox{0.9}{
\begin{tabular}{@{}lrll@{}}
\toprule
\textbf{Models}     & \textbf{Latency} & \textbf{Frames} & \textbf{Platform} \\ \cmidrule{1-4}
MLIDS & 275\,ms & per CAN frame & GTX Titan X \\
NovelADS & 128.7\,ms & 100 CAN frames & Jetson Nano \\
GIDS & 5.89\,ms & 64 CAN frames &  GTX 1080  \\
DCNN &  5\,ms & 29 CAN frames & Tesla K80 \\
TCAN-IDS & 3.4\,ms & 64 CAN frames & Jetson AGX \\
MTH-IDS~\cite{yang2021mth} &  0.574\,ms & per CAN frame & Raspberry Pi 3     \\
\textbf{MA-QCNN (DPU)} & 0.43\,ms & per CAN frame & Zynq FPGA    \\ 
\bottomrule
\end{tabular}}
\label{table:latcomp}\vspace{-10pt}
\end{table}

\begin{table}[t!]
\centering
\vspace{0.3cm}
\caption{Resource utilization breakdown for PL Accelerator of our proposed CAN IDS (XCZU3EG).}
\scalebox{1}{
\begin{tabular}{lrrrr}
\toprule
\textbf{Node} & \textbf{LUT} & \textbf{FF} & \textbf{BRAM} & \textbf{DSP} \\
\midrule
DPU           & 30726        &  48400           & 123            & 164  \\
\midrule
Overall &  32648 & 51347  & 123  & 164 \\
(\% usage) & (46.2\%) & (36.3\%) & (56.9\%) & (45.5\%) \\
\bottomrule
\end{tabular}}
\label{table:resourceutilization}
\vspace{-10pt}
\end{table}

\section{Conclusion}\label{concl}
In this paper, we present a lightweight multi-attack machine learning model integrated as an IDS accelerator on a hybrid FPGA-based ECU architecture that can successfully detect multiple attack models on a CAN bus.
The proposed lightweight model has comparable detection performance (almost identical or within 1\%) to the state of the art complex machine learning models for detecting DoS and fuzzing attacks, almost all of which require dedicated GPU acceleration or dedicated IDS-ECUs for line-rate detection. 
Our model also achieves a 25\% reduction in latency of processing per message while consuming less than 2\,W of active power when executing the model.
We believe that the proposed learning model and integration approach can be adapted for emerging in-vehicle networks like Automotive Ethernet and also for enabling a distributed intrusion detection framework for vehicular networks.
\section{Acknowledgement}
This research was supported by grants from NVIDIA and utilised NVIDIA RTX A6000 GPU.
\bibliography{references}
\bibliographystyle{ieeetr}

\end{document}